# Bin Packing Problem: A Linear Constant-Space $\frac{3}{2}$-Approximation Algorithm

*Abdolahad Noori Zehmakan\**

*Department of Computer Science, ETH University, Zurich, Switzerland, Abdolahad.noori@inf.ethz.ch*

## Abstract

Since the Bin Packing Problem (BPP) is one of the main *NP-hard* problems, a lot of approximation algorithms have been suggested for it. It has been proven that the best algorithm for BPP has the approximation ratio of $\frac{3}{2}$ and the time order of $O(n)$, unless $P = NP$. In the current paper, a linear $\frac{3}{2}$-approximation algorithm is presented. The suggested algorithm not only has the best possible theoretical factors, approximation ratio, space order, and time order, but also outperforms the other approximation algorithms according to the experimental results; therefore, we are able to draw the conclusion that the algorithms is the best approximation algorithm which has been presented for the problem until now.

## 1. Introduction

Similar to other optimization problems, the BPP is included among NP-hard problems. In recent years, several attempts have been made to solve this problem with the approximation approaches [1]. The problem comprises many real-world applications; for instance stock-cutting, loading trucks, railway carriages, and etc. As this problem is NP-hard, a lot of approximation algorithms have been presented for it. In this paper, we also present an approximation algorithm which performs so efficiently, both in practice and in theory.

In the Bin Packing Problem, n items were given with specified weights $a_1, a_2, \dots, a_n$ such that:

---

$$0 < a_1 \leq a_2 \leq \cdots \leq a_n < 1$$

With the purpose of packing the objects in the minimum number of bins and ensuring that the sum of the objects' weights does not exceed the bin capacity in any bin.

The BPP is similar to a decision problem called the Partition Problem. In the Partition Problem, we are given n positive integers $b_1$, $b_2$, ..., $b_n$ whose sum B=$\sum_{i=1}^{n} b_i$ is even and the purpose is to know if the partition of the set of indices {1, 2, ..., n} into sets S and T such that $\sum_{i \in S} b_i = \sum_{i \in T} b_i$, is possible. The partition problem is well-known to be *NP-complete*.

In recent years, different versions of the problem like two and three-dimensional BPP [2, 3], fragile objects [4], extendable bins [5], and controllable item sizes [6] have been discussed. On the other hand, the basic problem consists of two main versions named on-line and off-line. In the former, items are appeared without any knowledge about the sequence of them (neither the size nor the number of the items). In the latter, all of the items are given at the initial part of problem solving.

Simchi-Levi [7] proved that FFD (First Fit Decreasing), and BFD (Best Fit Decreasing) algorithms have an absolute worst-case ratio of $\frac{3}{2}$. These algorithms' time order is $O(n \log n)$. Zhang and Xiaoqiang [1] provided a linear time constant-space off-line approximation algorithm with absolute worst-case ratio of $\frac{3}{2}$. Moreover, they presented a linear time constant-space on-line algorithm and proved that the absolute worst-case ratio is $\frac{3}{2}$. In 2003, Rudolf and Florian [8] also presented an approximation algorithm for the BPP which has a linear running time and absolute approximation factor of $\frac{3}{2}$. It has been proven that the best algorithm for BPP has the approximation ratio of $\frac{3}{2}$ and the time order of $O(n)$, unless $P = NP$[7]. Noori Zehmakan [9] presents two approximation algorithms for this problem. The first one is a $\frac{3}{2}$-approximation algorithm, and the second one is a modified (linear time) version of the FFD algorithm.

As mentioned, different approximation algorithms [10, 11] have been presented for the BPP because of its importance. In this paper, a new creative algorithm with the best possible factors, time order, space order, and approximation ratio is presented.

Then, the proposed algorithm is compared with the approximation algorithms with the best possible approximation factor and time order ($\frac{3}{2}, O(n)$).

The reminder of the paper is divided into two sections. In the first section, the suggested algorithm is presented and in the second one, the computational results are demonstrated and discussed.

## 2. The Proposed Algorithm

The inputs are separated into two different groups. The first one contains *Small items* or *S items* (the items that are not larger than 0.5) and the *Large items* or *L items* (the items that are larger than 0.5). The main idea of the algorithm is to put the inputs into ten equal ranges and try to match the items in complementary ranges.

After classification items into ten equal ranges, the algorithm matches L items with S items and creates new items and continues this procedure until finishing L items. Finally, it joins remaining S items to each other.

---

***Function $F_1$*** $((x, y))$;

*Choose an item $a$ in $(x, y)$;*

***If*** $((1 - y, 1 - x)$ *is not empty*)
  *Choose an item $b$ in $(1 - y, 1 - x)$ at random & let $c = a + b$;*
  ***If*** $(c \leq 1)$
    *Put $c$ in the appropriate range and remove $a, b$;*
    ***Return*** *0;*
  ***Else Return*** *$a$;*
***Else Return*** *$a$;*

---

***Function $F_2$*** $(a, (x, y))$
***If*** $((x, y)$ *is not empty*)
  *Choose an item $b$ in $(x, y)$ at random & let $c = a + b$;*
  *Put $c$ in the appropriate range and remove $a, b$;*
  ***Return*** *1;*

```
Function F₃ ()
For ((x, y) = (0.5, 0.6); x < 1.0, ; (x = x + 0.1, y = y + 0.1))
  While ((x, y) is not empty)
    a=( F1( (x, y))
    If ( a = 0)
      Continue;
    Else
      f=0 & i=1;
      while ( g=0 & 1-y-i*0.1>0.0)
        g=F2 (a, (1-y- i*0.1, 1-x-i*0.1)) & i++;
      If( 1-y-i*0.1<0.0)
        bin-counter ++;
        remove a;
```

## Algorithm:

```
Read n inputs & classify them into 10 ranges (0, 0.1), (0.1, 0.2) ... (0.9, 1);
F3 ()
For ((x, y) =(0.4,0.5); x ≥0.0; (x=x-0.1, y=y-0.1))
  While ((x, y) is not empty)
    Choose two items a, b in (x, y) & let c = a + c;
    Put c in the appropriate range and remove a, b;
    F3 ();
```

In addition to the mentioned pseudo code, for the ease of comprehension, a different version of the algorithm is also presented.

1- **Read** n inputs and **classify** them into 10 equal ranges (0.0, 0.1),…, (0.9, 1.0).

2- If there are no items in (0.5, 0.6) go to **line 8**.

3- Choose an item **a** in (0.5, 0.6). If there is at least an item in (0.4, 0.5), choose an item **b** in it randomly, and set **c** = **a** + **b**. If **c** ≤ **1**, put it in the appropriate range as a new item and **remove a** and **b** and go to **line 2**.

4- If there is at least an item in (0.3, 0.4), choose an item **b** in it at random, and set **c** = **a** + **b,** put it in the appropriate range as a new item, **remove a** and **b** and go to **line 2**.

5- If there is at least one item in $(0.2, 0.3)$, choose an item *b* in it randomly and set $c = a + b$, and put it in the appropriate range as a new item, **remove** *a* and *b*, and go to **line 2**.

6- If there is at least one item in $(0.1, 0.2)$, choose an item *b* in it at random and set $c = a + b$, put it in the appropriate range as a new item, **remove** *a* and *b*, and go to **line 2**.

7- If there is at least one item in $(0, 0.1)$, choose an item *b* in it randomly and set $c = a + b$. If $c \leq 1$, put it in the appropriate range as a new item and **remove** *a* and *b* and go to **line 2**.

8- If there are not any items in $(0.6, 0.7)$ go to **line 9** else do steps **4 to 7** for this range.

9- If there are not any items in $(0.7, 0.8)$ go to **line 10** else do steps **5 to 7** for this range.

10- If there are not any items in $(0.8, 0.9)$ go to **line 11** else do steps **6 and 7** to the range.

11- If there are not any items in $(0.9, 1)$ go to **line 12** else do step **7** for this range.

12- If there are not any items in $(0.4, 0.5)$ go to **line 14**.

13- Choose two items *a* and *b* in $(0.4, 0.5)$ at random. Set $c = a + b$. Put it in the appropriate range as a new item and **remove** *a* and *b* and go to **line 2**.

14- If there are not any items in $(0.3, 0.4)$ go to **line 16**.

15- Choose two items *a* and *b* in $(0.3, 0.4)$ randomly. Set $c = a + b$. Put it in the appropriate range as a new item and **remove** *a* and *b* and go to **line 2**.

16- If there are not any items in $(0.2, 0.3)$ go to **line 18**.

17- Choose two items *a* and *b* in $(0.2, 0.3)$ at random. Set $c = a + b$. Put it in the appropriate range as a new item and **remove** *a* and *b* and go to **line 2**.

18- If there are not any items in $(0.1, 0.2)$ go to **line 20**.

19- Choose two items *a* and *b* in $(0.1, 0.2)$ randomly. Set $c = a + b$. Put it in the appropriate range as a new item and remove *a* and *b* and go to **line 12**.

20- If there are not any items in $(0, 0.1)$ go to **line 22**.

21- Choose two items *a* and *b* in $(0, 0.1)$ randomly. Set $c = a + b$. Put it in the appropriate range as a new item and remove *a* and *b* and go to **line 18.**

22- **End**.

The state which there is only one $S$ item in the range taken into consideration has been ignored in steps 13, 15, 17, 19 and 21 for the ease of algorithm's comprehension, but definitely, it cannot be ignored in the complete form.

The algorithm attempts to match each $L$ item with the best possible $S$ item and creates bins with the least free space. Obviously, the proposed algorithm's time order is $O(n)$ in that in each step, the algorithm eliminates two items and creates a new one, and the process of each step is done in at most k time unit. Thus, its time order equals to $O(kn)$, and k is a constant; therefore, the suggested algorithm is a linear time one.

We have proved that the approximation ratio of the algorithm is $\frac{3}{2}$, and interested reader can find the exact and complete proof in [12]. Actually, since the proof is so long, we have presented this proof as a distinct document on arXiv. We present a sketch of the proof in this paper, and the complete and detailed proof is disregarded here inasmuch as the main purpose of this article is discussing and comparing the results of the suggested algorithm with other algorithms, both in theory and in practice.

***Definition 1***: An output bin is named an *L pack* if it contains one $L$ item, and it is called an *S pack* if it has no L items.

***Definition 2***: $P$ is the number of output bins in OPT solution, and $P^*$ is the number of output bins in the proposed algorithm.

Two possible kinds of output bins ($L$ pack and $S$ pack) are shown in figure 1.

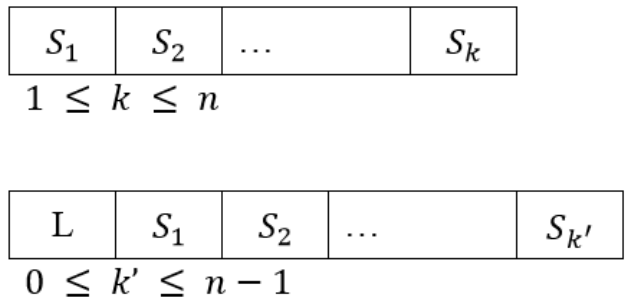

*Figure 1: The states of the output bins in a OPT solution*

***Lemma 1***: If at least $\frac{2}{3}$ size of each output bin is full, $\frac{P^*}{P} \leq \frac{3}{2}$.

***Proof***: Consider the worst state that all output bins are completely full in OPT solution, and suppose that $W$ is the sum of size of input items. Then,

$$P \geq W \ \& \ P^* \leq \frac{W}{\frac{2}{3}} \Rightarrow P^*/P \leq \frac{3}{2} \ \blacksquare$$

***Lemma 2***: If all inputs are $S$ items, the ratio of $P^*$ to $P$ is not more than $\frac{3}{2}$.

***Proof***: Based on the algorithm, items in $(0.4, 0.5)$ will be matched with each other two by two and create new items between 0.8 and 1.0. Then, these new items try to be matched with item in $(0, 0.1)$ or $(0.1, 0.2)$. Afterward, each final created item will be put in one bin. Therefore, at least 0.8 sizes of each output bin in this step is full.

After finishing the items in $(0.4, 0.5)$, items in $(0.3, 0.4)$ are joined together in binary and triad groups. This range will be discussed at the end of this proof precisely.

Definitely, after the items in $(0.3, 0.4)$, the items relevant to $(0.2, 0.3)$ will be matched with each other. Obviously, in this condition, there is no items in $(0.3, 1.0)$. Two items from $(0.2, 0.3)$ are matched with each other and create a new item between 0.4 and 0.6. After that, the new created item is matched with an item in $(0.2, 0.3)$ and will make a new item. This process continues up to a point that the created item becomes larger and larger. Finally, at least 0.7 of each output bin's size is full because the largest item of Small set in this step is 0.3.

Certainly, after finishing the items in $(0.2, 0.3)$, no items exist between 0.2 and 1.0. Obviously, with the mentioned logic for $(0.2, 0.3)$ in this step also at least 0.8 size of each output bin is full because the largest item of Small set is 0.2. In the next step, the remaining items are in $(0, 0.1)$. Definitely, in this step at least 0.9 of each output bin's size is full because the largest item of Small set is 0.1.

According to *lemma 1*, ranges $(0, 0.1)$, $(0.1, 0.2)$, $(0.2, 0.3)$ and $(0.4, 0.5)$ do not contradict the claim that $\frac{P^*}{P}$ is not larger than $\frac{3}{2}$, on the other hand $(0.3, 0.4)$ can cause some problems because this range is able to create bins that are smaller than $\frac{2}{3}$. In the worst-case, all items are in $(0.3, 0.4)$ because the total free space in output bins and the number of output bins in the algorithm will be maximized. Even in this condition, $Max\left(\frac{P^*}{P}\right) = \frac{3}{2}$ because:

$$\text{Assume } n = 3m \Rightarrow P^* \leq \frac{3m}{2}, P \geq m \Rightarrow Max\left(\frac{P^*}{P}\right) \leq \frac{3}{2} \blacksquare$$

**Theorem 1:** The approximation ratio of the suggested algorithm is $\frac{3}{2}$.

**Sketch of proof:** First, four types of errors (error here means the deviation from OPT solution) which are all possible types of the algorithm's errors are introduced, and if no errors exist, then the algorithm will result in optimal solution. Next, the algorithm with only error type1 is discussed and it is proved that the approximation ratio in this condition is $\frac{3}{2}$. Afterwards, other types of errors are added one by one. Finally, it is proved that if all types of the errors exist, the approximation ratio will be $\frac{3}{2}$, again. Note that in all steps we consider the worst condition. Obviously, If $\frac{P^*}{P}$ is never more than $\frac{3}{2}$, the approximation ratio is $\frac{3}{2}$.

*The algorithm's possible errors*:

1- An L item is matched with an S item which belongs to an S pack.

2- An L item is matched with an S item which belongs to an L pack such that the L item relevant to the L pack and the first L item belong to the same range.

3- An L item is matched with an S item which belongs to an L pack such that the L item in this pack belongs to a range except the range which the first L item is from.

4- Error in matching S items (both in S packs and L packs): S items are matched in a wrong manner; it means that they are matched in a different way from the optimal solution.

Based on the algorithm an $L$ item begins to pick up an $S$ item. The $L$ item can make some possible errors and pick up an $S$ item which belongs to an $L$ pack such that the $L$ item of the $L$ pack and the mentioned $L$ item belong to the same range (error type 2) or not (error type 3), It may pick up an $S$ item from an $S$ pack (error type 1). Finally, $S$ items can be matched with each other in a wrong way (error type 4). The attempts will be made to maximize the errors in each step of the proof to engender the worst condition. And it is observed in all states approximation ratio is $\frac{3}{2}$.

In some condition that depend on errors type 1, 2 and 3, it is shown $\frac{2}{3}$ space of all output bins are full, and based on the *lemma 1* $\frac{P^*}{P} \leq \frac{3}{2}$. Moreover, in the situations that error type 4 happens based on *lemma 2* $\frac{P^*}{P} \leq \frac{3}{2}$, again. Finally, by considering all types of errors together, it is proved that the approximation ratio of the algorithm is $\frac{3}{2}$. ∎

## 3. Computational experiments

In this section, the suggested algorithm are compared with three other approximation algorithms which enjoy the best possible theoretical criteria, approximation ration, time order, and space order . This comparison has been drawn based on the eight standard data sets from **OR-LIBRARY** [13].

For the comparison, we need a parameter which is strongly relevant to approximation ratio. The ***Ratio*** is defined [9] as the proportion of the proposed algorithm's solution to the OPT solution. Obviously, *Ratio* has a direct relationship with approximation ratio. Thus, *Ratio* can be utilized as a factor for measuring approximation algorithms' performances.

As mentioned, the standard data sets in OR-LIBRARY are used for simulations. Each data set contains 20 input instances such that each input instance includes a bunch of items which are a standard input for the Bin Packing Problem. Based on the eight instances the suggested algorithm has been compared with the *Guochuan's* [1] and the *Berghammer*'s algorithms [8]. The results of these comparisons for $bp_i$ $1 \leq i \leq 8$ are shown in figures $j$ $2 \leq j \leq 9$, respectively.

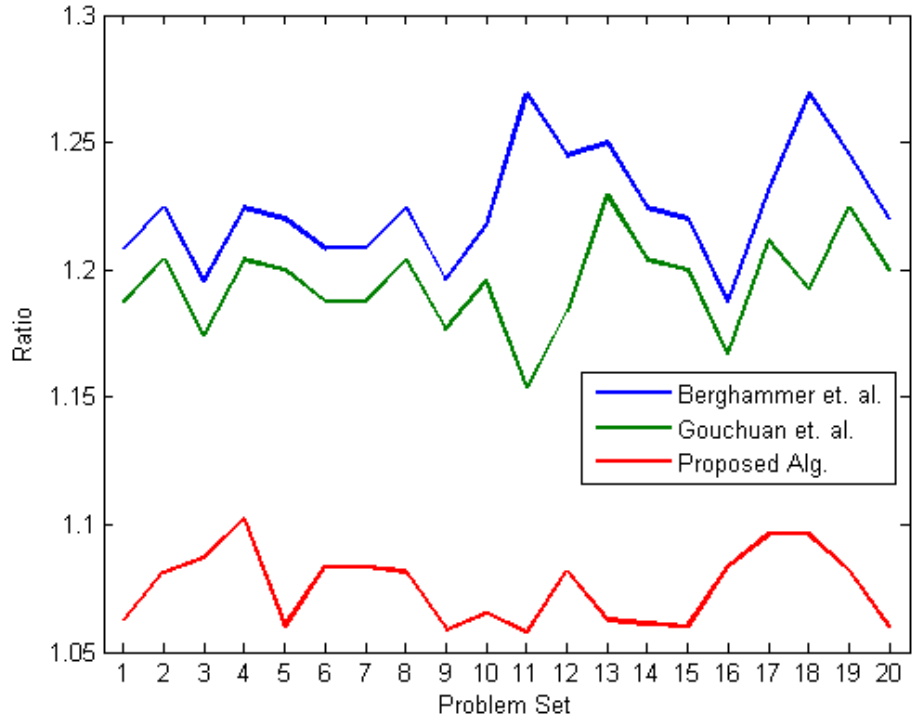

*Figure 2*: *The comparison of Ratio of the algorithms for 20 input instances in data set bp1*

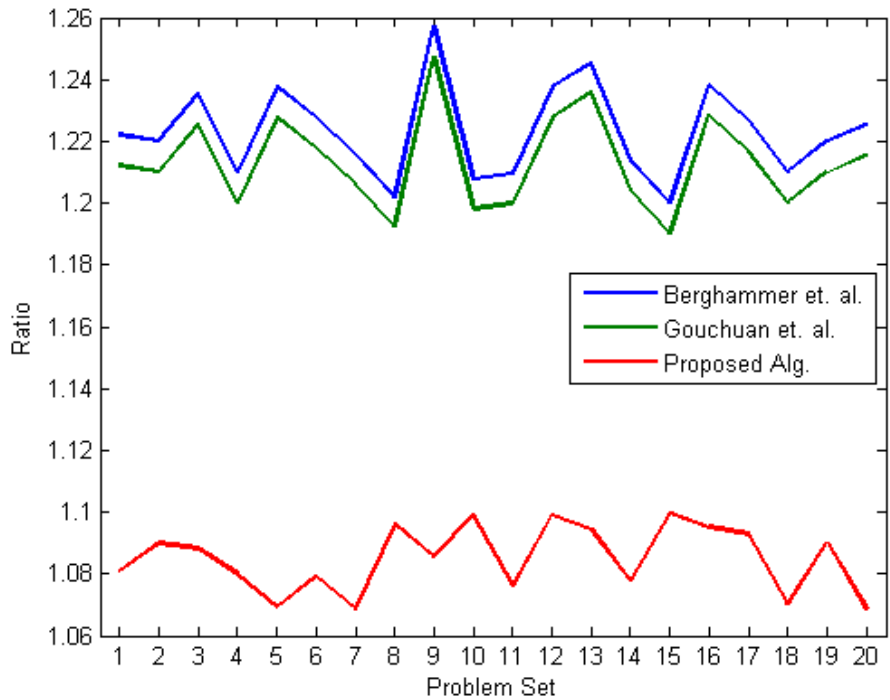

*Figure 3*: *The comparison of Ratio of the algorithms for 20 input instances in data set bp2*

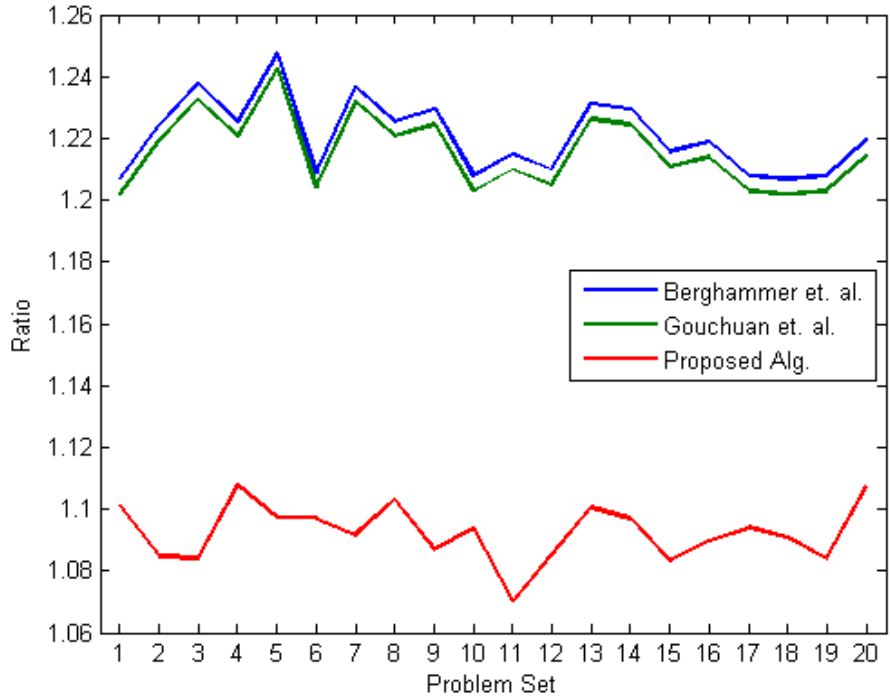

*Figure 4*: *The comparison of Ratio of the algorithms for 20 input instances in data set bp3*

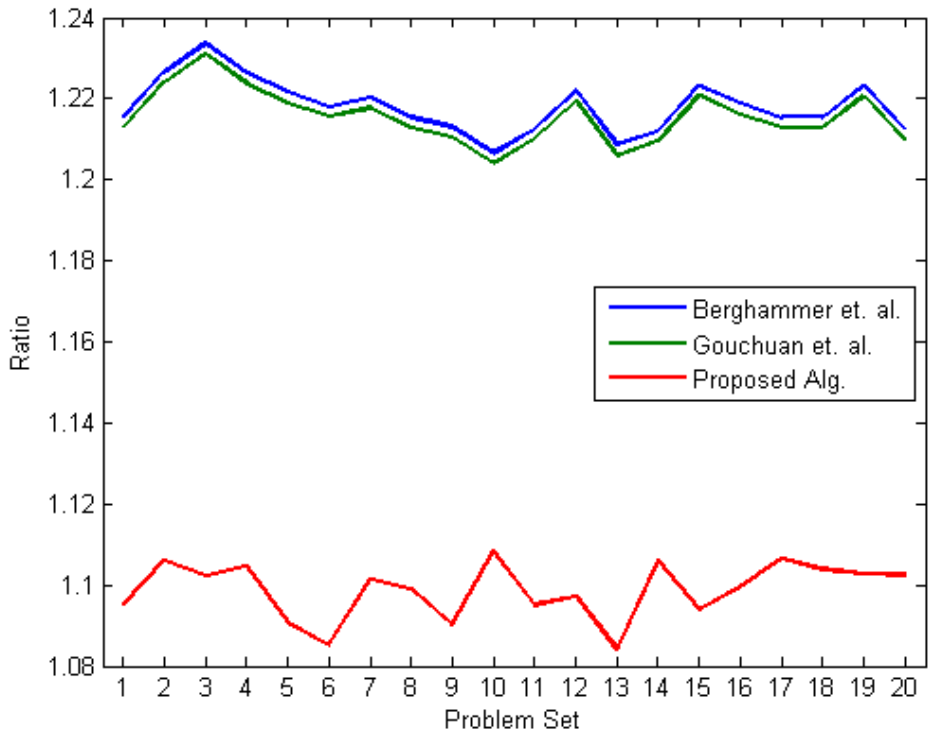

*Figure 5*: *The comparison of Ratio of the algorithms for 20 input instances in data set bp4*

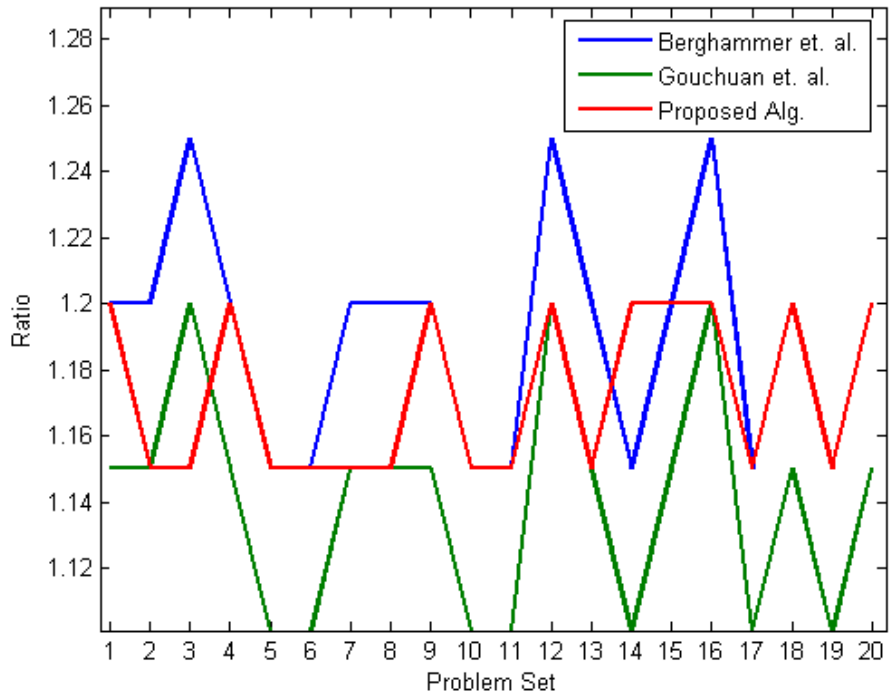

*Figure 6*: *The comparison of Ratio of the algorithms for 20 input instances in data set bp5*

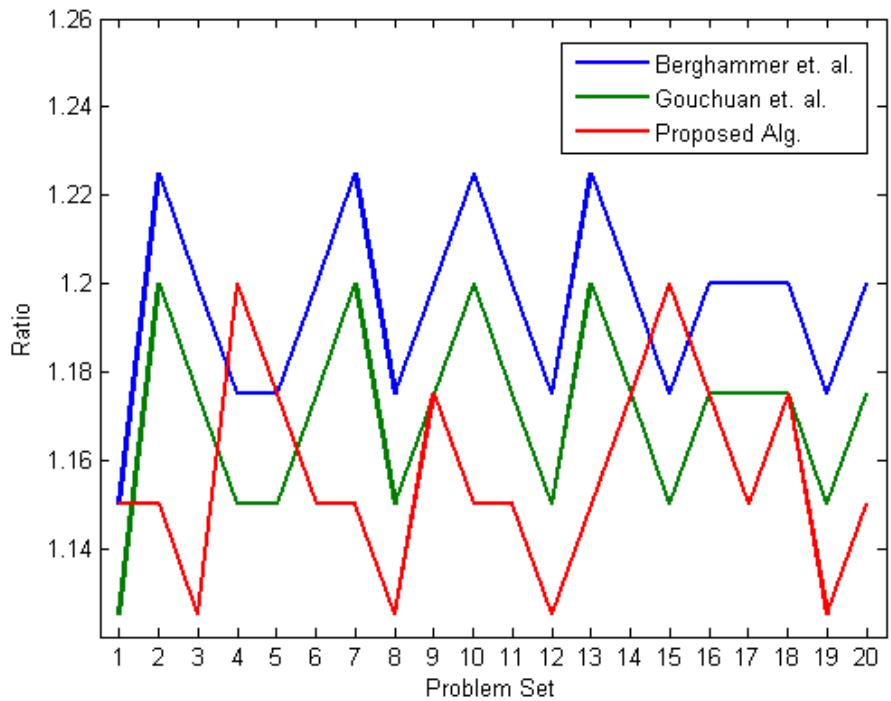

*Figure 7*: *The comparison of Ratio of the algorithms for 20 input instances in data set bp6*

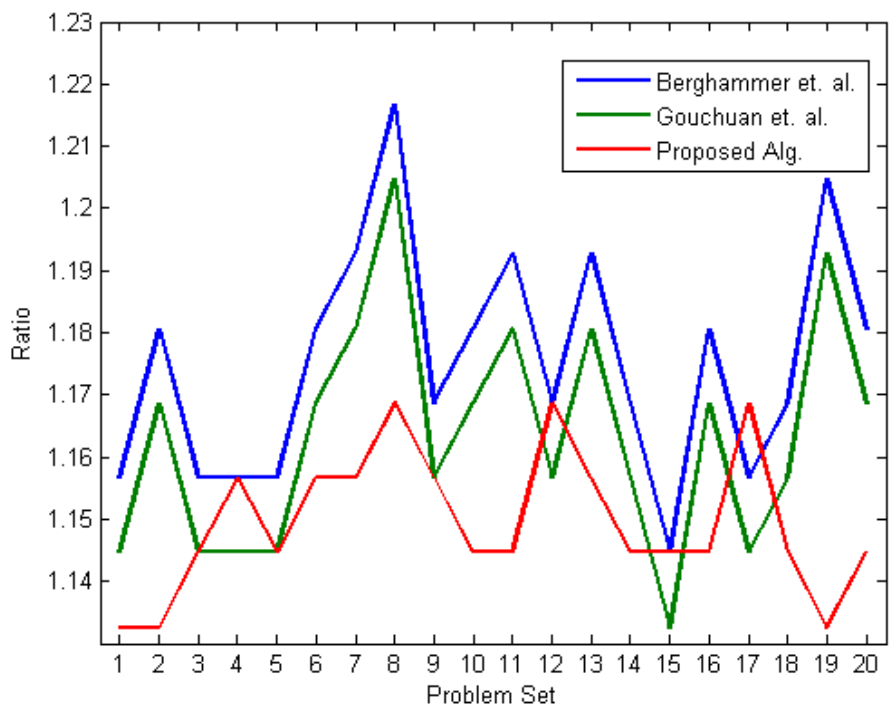

*Figure 8*: *The comparison of Ratio of the algorithms for 20 input instances in data set bp7*

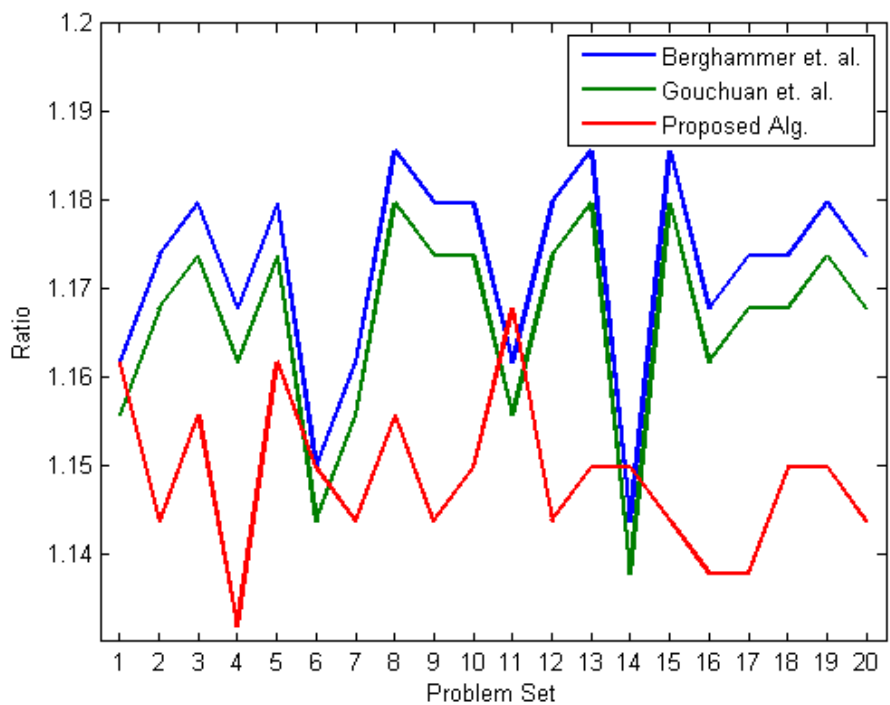

*Figure 9*: *The comparison of Ratio of the algorithms for 20 input instances in data set bp8*

It is obvious that the performance of the proposed algorithm is much more acceptable than the two other ones. Note that the mentioned approximation algorithms are only ones which have the best possible approximation factor and running time. In each figure, the results are computed for 20 input instances relevant to the corresponding $bp_i$, but for the ease of understanding the points relevant to each algorithm are joined by a line in each figure.

The averages of the results for the eight data sets by the approximation algorithms are shown in figure 10. Obviously, the suggested algorithm performs much better than other ones in average. Figure 10 shows that the proposed algorithm except in data set *bp5* holds smaller ratios. Furthermore, it also shows that the performance of *Guochuan's* algorithm is more acceptable than the performance of *Berghammer*'s algorithm.

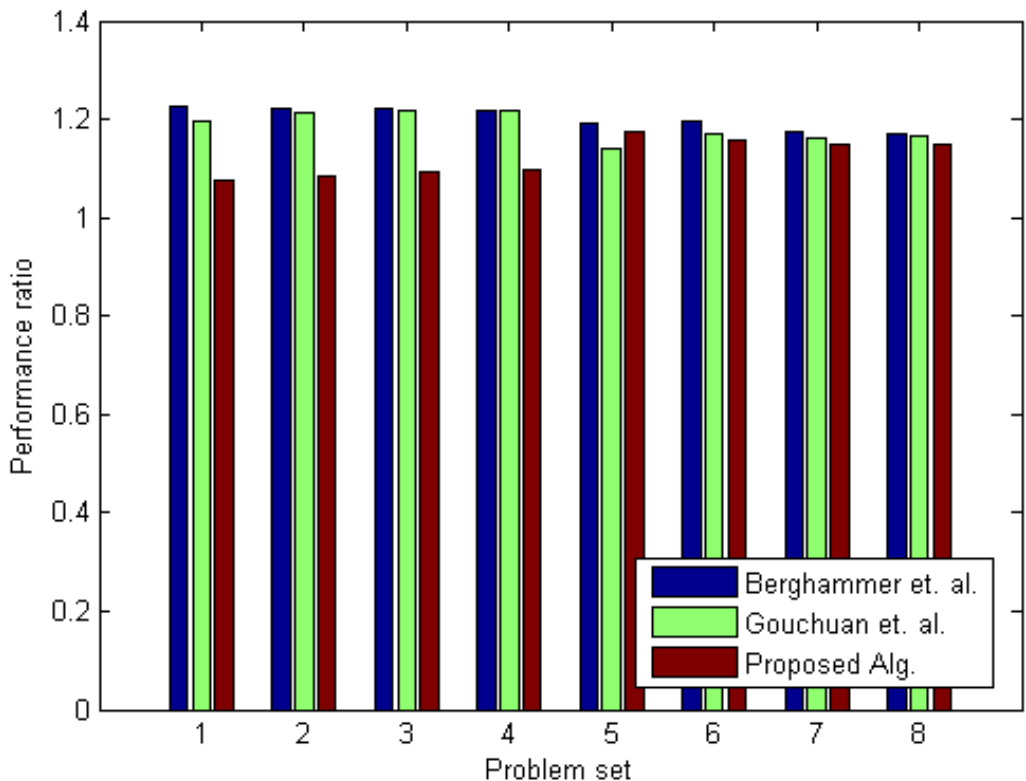

***Figure 10***: *The average of the algorithms' ratios for the eight data sets.*

Figure 11 shows the results of comparing the suggested algorithm with FFD algorithm. The suggested algorithm has similar performances with FFD in instances *bp4*, *bp6, bp7*, and *bp8* (The input instances whose number of items are much more than others), but its performance is a little worse than FFD algorithm in *bp1*, *bp2*, *bp3*, and *bp4*. Note that FFD algorithm's time order is $O(nlogn)$ in average and $O(n^2)$ in worst-case, but the proposed algorithm is a linear time algorithm. Furthermore, FFD is an online-space algorithm, but the proposed algorithm is a constant-space one. It means that FFD saves all bins until the last step. Then, its time order and space order is much greater than the suggested algorithm. Therefore, the suggested algorithm performs much better than FFD, especially in big data-sets inasmuch as its time order and space order is considerably better than FFD and enjoys more acceptable Ratios in big data-sets.

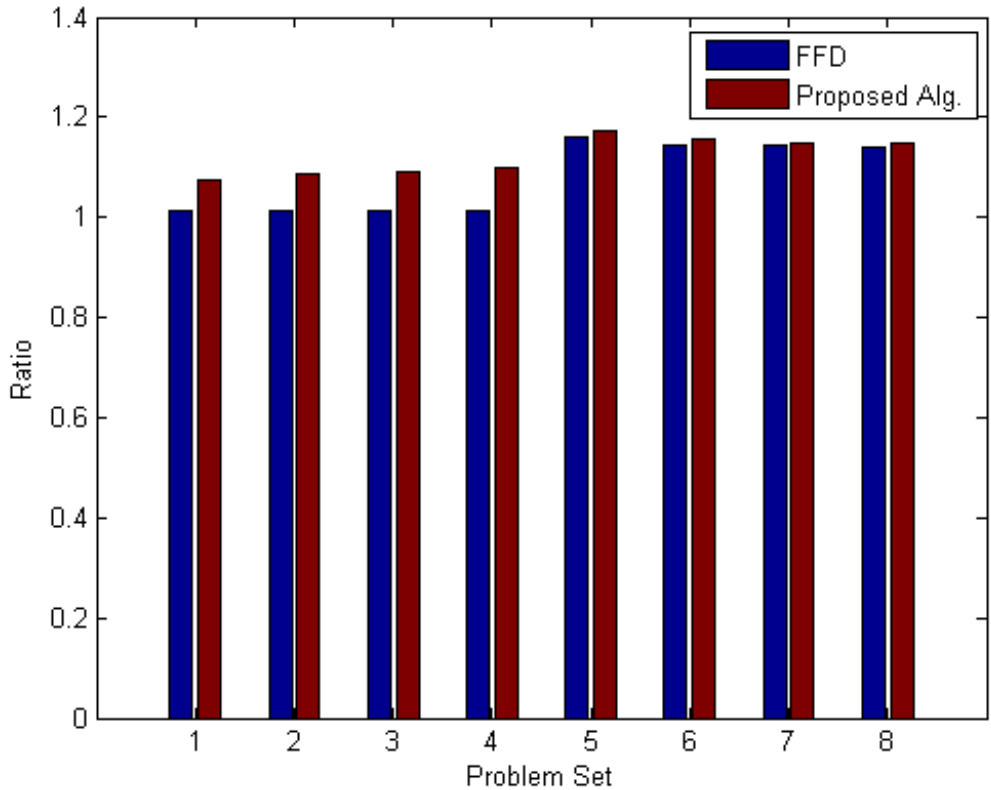

*Figure 11*: *The average of the suggested algorithm' ratios and FFD algorithm's ratios for the eight data sets.*

Furthermore, this algorithm is also much considerably better than our previous algorithm [9] inasmuch as this algorithm not only is linear time, but also performs much better in practice based on the comparison of the experimental results.

In recapitulation, we can claim that the presented algorithm is the best approximation algorithm for the BPP up to now in theory and also in practice.

## 4. The Enhancement of The Algorithm

In this section two ideas which can help the algorithm to be enhanced are discussed.

### 4.1. Creating more ranges

In the suggested algorithm, we consider ten ranges, but obviously, we can increase them in number. For instance, see the ranges in figure 12.

| 0.5-0.55 | 0.55-0.6 | 0.6-0.65 | 0.65-0.7 | 0.7-0.75 | 0.75-0.8 | 0.8-0.85 | 0.85-0.9 | 0.9-0.95 | 0.95-1 |
|---|---|---|---|---|---|---|---|---|---|
| 0.45-0.5 | 0.4-0.45 | 0.35-0.4 | 0.3-0.35 | 0.25-0.3 | 0.2-0.25 | 0.15-0.2 | 0.1-0.15 | 0.05-0.1 | 0-0.05 |

*Figure 11: Another possible ranging (using 20 equal ranges)*

The algorithm can be executed with the new ranging, and it seems that this new algorithm must be more efficient. For example, if the inputs are 0.41, 0.47, 0.48, 0.59, 0.53 and 0.52, the algorithm utilizes 3 or 4 bins to pack all items in both states, but the probability of exploiting just 3 bins for the algorithm in state of new ranging is higher.

This procedure can be continued, and each range can be divided into two new equal ranges, which will subsequently lead to $20, 40, 80, ...$ ranges. The authors claim that by using the new ranging the approximation ratio of the algorithm is still also 3/2, and its time order is linear too. Obviously, the time order of the algorithm remains $O(n)$ in this state. Moreover, the proof relevant to approximation ratio can be generalized for the state of new ranging in that the ranging does not impose any special restrictions on the proof.

Certainly, if the number of ranges is increased, the efficiency of the algorithm will be improved because input items will be sorted more accurately. On the other hand, it is noticeable that the number of ranges cannot depend on the number of inputs because in such case the time order of the algorithm would not be linear anymore. Some constant numbers can be considered, and one of them can be selected based on the number of inputs. For example, consider four possible ranging (10 equal ranges, 80 equal ranges, 640 equal ranges, and 5120 equal ranges). Definitely, two good choices for 20 and 390000 inputs can be 10 and 5120 equal ranges, respectively.

Therefore, it seems the algorithm by using this ranging method (for example, 20 ranges) usually performs much better than the primary version. It is known that its performance is not worse than the original one because it is a special state of the main algorithm. On the other hand, input items are more sorted, and definitely, the new algorithm will perform more efficiently. For example, consider three inputs $x$, $y$ and $z$ that are in $(0.4, 0.6)$.

| 0.5-0.55 (x) | 0.55-0.6 |
|---|---|
| 0.45-0.5 (y) | 0.4-0.45 (z) |

The algorithm with 20 ranges will match $x$ and $y$ with each other. Now, if $x$ and $y$ are matched with each other in OPT solution too, this method performs efficiently. If $x$ and $z$ are matched in OPT solution, this method performs efficiently again because $z$ can be replaced by $y$ ($z \leq y$). On the other hand, the main algorithm will match $x$ with $y$ or $z$. If $x$ and $y$ are matched with each other in OPT solution and $x$ & $z$ are matched in the main algorithm, the algorithm will not perform optimally. In conclusion, the algorithm which uses the suggested ranging method usually acts more efficiently than the main algorithm.

### 4.2. Scaling method

Another idea that might make the algorithm more efficient is scaling. Based on the algorithm, after finishing L items, S items are matched with each until finishing all input items. After finishing L items, remaining items are between 0.0 and 0.5, but these ranges can be scaled to the primary ranges (0.0 to 1.0). After that, some new L items will be created (We call them virtual L items). These new items are matched with each other based on the suggested algorithm. This procedure continues until all items are processed, meaning the scaling can be repeated several times. These virtual bins (bins that are created through scaling) must be divided by $2^i$ to create main items ($i$ is the number of times which we have done scaling until that step). Finally, after dividing, the output bins will be matched $2^i$ by $2^i$ with each other and create final output bins. Note, it is necessary to specify a boundary for the level of scaling unless this procedure can continue lots of times.

For instance, suppose that we are given the following inputs.

*Primary configuration: 0.4, 0.3, 0.3, 0.2, 0.2, 0.1*

Based on the scaling method, these items will be scaled, and we have the following virtual configuration.

*Virtual configuration: 0.8, 0.6, 0.6, 0.4, 0.4, 0.2*

According to the scaling method there will be the following virtual bins. Note $i = 1$ in this step because we have just done scaling once.

*Virtual Output Bins: (0.6, 0.4), (0.6, 0.4), (0.8, 0.2)*

Now, it is enough to divide all virtual bins by 2 (actually $2^i$ such that $i = 1$) to create new items. After dividing, we will have the following items:

*Output items after dividing: 0.5, 0.5, 0.5*

Now, these item are matched 2 by 2 (actually $2^i$ such that $i = 1$), and then we have:

*Bin 1 = (0.5, 0.5) = (0.4, 0.1, 0.3, 0.2)*

*Bin 2 = (0.5) = (0.3, 0.2)*

Therefore, we need two bins to pack the items.

## 5. Conclusion

As we mentioned, BPP is a significant problem which is exploited in various fields and different applications. Because of its NP-hard nature, lots of approximation algorithms have been presented for it. In this paper, an approximation algorithm with the best possible theoretical factors, linear time order, constant-space, and approximation ratio of 3/2, was presented. The algorithm works based on dividing input items into ten equal ranges, and matching them with each other in a special way.

Furthermore, the proposed algorithm, based on a standard library, were compared with three other approximation algorithms which are the only algorithms which enjoy the best possible theoretical factors. The results showed that the suggested algorithms considerably outperforms other approximation algorithms previously presented for this problem in practice too.

Finally, we suggested two constructive ideas which could enhance the algorithm.